\documentclass[journal=jacsat,manuscript=article]{achemso}

\usepackage{graphicx}
\usepackage{color}
\usepackage[dvipsnames]{xcolor}
\usepackage{float}
\usepackage{amsmath,amssymb}
\usepackage{soul}
\usepackage{float}

\title{Epitaxial growth of SiC on (100) Diamond}

\author{A. Tsai}
\affiliation{School of Physics, University of Melbourne, Parkville, VIC 3010, Australia}

\author{A. Aghajamali}
\affiliation{Department of Physics and Astronomy, Curtin University, Perth, Western Australia 6102, Australia}

\author{N. Dontschuk}
\affiliation{School of Physics, University of Melbourne, Parkville, VIC 3010, Australia}
\altaffiliation{Centre for Quantum Computation and Communication Technology, School of Physics, University of Melbourne, Parkville, VIC 3010, Australia}

\author{B. C. Johnson}
\affiliation{Centre for Quantum Computation and Communication Technology, School of Physics, University of Melbourne, Parkville, VIC 3010, Australia}

\author{M. Usman}
\affiliation{Centre for Quantum Computation and Communication Technology, School of Physics, University of Melbourne, Parkville, VIC 3010, Australia}

\author{A. K. Schenk}
\affiliation{Department of Chemistry and Physics, La Trobe Institute for Molecular Science, La Trobe University,
Victoria 3086, Australia}

\author{M. Sear}
\affiliation{Department of Chemistry and Physics, La Trobe Institute for Molecular Science, La Trobe University,
Victoria 3086, Australia}

\author{C. I. Pakes}
\affiliation{Department of Chemistry and Physics, La Trobe Institute for Molecular Science, La Trobe University,
Victoria 3086, Australia}

\author{L. C. L. Hollenberg}
\affiliation{Centre for Quantum Computation and Communication Technology, School of Physics, University of Melbourne, Parkville, VIC 3010, Australia}

\author{J. C. McCallum}
\affiliation{School of Physics, University of Melbourne, Parkville, VIC 3010, Australia}

\author{S. Rubanov}
\affiliation{Bio21 Institute, University of Melbourne, Parkville, VIC 3010, Australia}

\author{A. Tadich}
\affiliation{Australian Synchrotron, 800 Blackburn Road, Clayton, Victoria 3168, Australia}

\author{N. A. Marks}
\affiliation{Department of Physics and Astronomy, Curtin University, Perth, Western Australia 6102, Australia}

\author{A. Stacey}
\affiliation{School of Science, RMIT University, Melbourne, VIC 3001, Australia}
\altaffiliation{Centre for Quantum Computation and Communication Technology, School of Physics, University of Melbourne, Parkville, VIC 3010, Australia}
\email{alastair.stacey@rmit.edu.au}

\begin{document}

\begin{abstract}
    We demonstrate locally coherent heteroepitaxial growth of silicon carbide (SiC) on diamond, a result contrary to current understanding of heterojunctions as the lattice mismatch exceeds $20\%$. High-resolution transmission electron microscopy (HRTEM) confirms the quality and atomic structure near the interface. Guided by molecular dynamics simulations, a theoretical model is proposed for the interface wherein the large lattice strain is alleviated via point dislocations in a two-dimensional plane without forming extended defects in three dimensions. The possibility of realising heterojunctions of technologically important materials such as SiC with diamond offers promising pathways for thermal management of high power electronics. At a fundamental level, the study redefines our understanding of heterostructure formation with large lattice mismatch. 
\end{abstract}


Diamond is a wide-bandgap semiconductor with an extreme thermal conductivity of 2400~W/m/K \cite{isberg2002high} that is of acute interest for novel quantum and high power electronic devices \cite{wort2008diamond,willander2006silicon,trew1991potential,okushi2001high}. It would be highly desirable to incorporate diamond into electronic structures such as high power silicon carbide (SiC) circuits to improve device performance; this is also true for nanoscale silicon device structures that are limited by high temperature \cite{meyer2003silicon, weitzel1996silicon}.
The small lattice parameter of single crystal diamond, however, limits the choice of materials with which it can form heterostructures. Typical problems include, amorphous growth or significant interfacial strain relief by defects such as screw dislocations, threading faults, and twinned growth \cite{chen1996structural, people1985calculation, bean1984ge, freund2004thin}. Such defects compromise the quality of the heteroepitaxial layer by extending through the grown material and can substantially degrade thermal transport and layer adhesion \cite{dong2014relative, cervenka2012diamond,imura2011demonstration, goyal2010reduced}. Furthermore, growth of single crystal diamond on different materials is prohibitively difficult and typically results in polycrystalline material \cite{jiang1995diamond,khmelnitskiy2015prospects,schreck2014large}. This study reports nearly strain-free epitaxial growth of SiC on diamond and paves the way for exploiting the promising properties of diamond in high-power and high-frequency electronics \cite{yoder1996wide}.

Heteroepitaxy with single crystal diamond growth is difficult and subjects the substrate material to the harsh diamond growth environment \cite{schreck2001diamond, schreck2014large} and also typically results in polycrystallinity \cite{koizumi2018power,aleksov2003diamond, stoner1992textured}. These considerations, together with the recent introduction of wafer-scale single crystal diamond \cite{yamada20142,yamada2013uniform,schreck2014large}, make it highly desirable to use diamond as the substrate upon which high-power electronic materials are then epitaxially grown. To date, only AlN and AlGaN/GaN heterostructures have successfully used diamond as a substrate for growth~\cite{hirama2010heterostructure, hirama2011algan, alomari2010algan}, though cases of diamond growth forming patches of coherence with 3C-SiC have been reported \cite{koizumi2018power} which were surrounded by unoriented and polycrystalline material with poor interface characteristics. Unfortunately, the mismatch-induced strain in the AlN and AlGaN/GaN structures is relieved by a high density of defects in the AlN lattice which also created a large amount of dangling bonds \cite{hirama2010heterostructure} and high electron mobility transistors made of AlGaN/GaN on diamond required thick buffer layers to construct operational devices \cite{alomari2010algan, hirama2011algan, hirama2012growth}.

\begin{figure*}
    \centering
    \includegraphics[width = 0.95\textwidth]{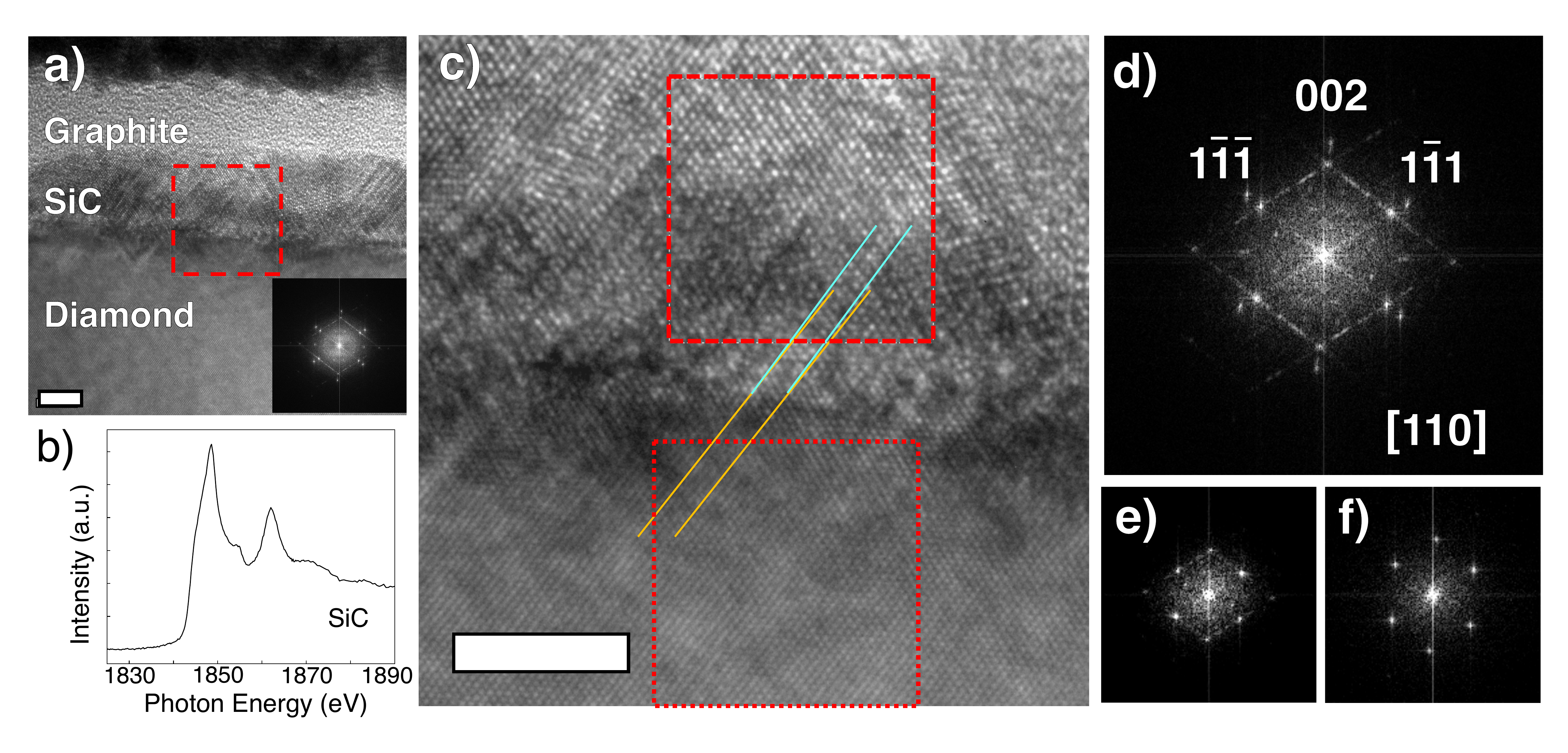}
    \caption{(a) Cross-sectional HRTEM (5~nm scale bar) of the prepared sample, showing graphite, SiC and diamond layers. The image is taken in the $[110] $ direction, and the inset shows the FFT of the entire image. (b) SiC NEXAFS of the surface after 1350$^\circ$C anneal. (c) Magnification of the selected area in (a) demonstrating a coherent interface between SiC and diamond; slight misorientation is shown via the orange and ligh blue lines. (d) is the FFT of (c) with clear and largely oriented SiC and diamond signals. (e) and (f) are FFTs of the dashed and dotted regions, respectively, highlighted in (c) that verify the crystal quality of SiC and diamond.}
    \label{figure1}
\end{figure*}

In this work we demonstrate that SiC can be grown heteroepitaxially on single crystal diamond, producing a coherent and seemingly strain-free interface in localized regions by utilizing Si surface termination  procedures developed in earlier work \cite{schenk2015formation}. The absence of measured strain here is remarkable given the approximate 22\% lattice mismatch of SiC relative to diamond and the high degree of coherence between the lattices suggests that conventional strain relief mechanisms are not present. Atomistic simulations are performed to understand the nature of the interface and suggest that the lattice mismatch is resolved through bonding reconstructions that are restricted to the interfacial layers without extending into the grown film. These reconstructions eliminate dangling bonds that otherwise exist at low-quality, highly-mismatched interfaces.


A high-resolution transmission electron microscopy (HRTEM) image of the fabricated interface is shown in Fig.~\ref{figure1} with accompanying Fast Fourier Transforms (FFTs) and a Near Edge X-ray Absorption Fine Structure (NEXAFS) spectrum. Fig.~\ref{figure1}a shows three distinct layers: a protective layer of graphite used for TEM handling, the grown SiC film with a cubic lattice structure, and the diamond substrate. The 3C-SiC layer is confirmed with NEXAFS in Fig.~\ref{figure1}b \cite{chang1999x,prado2003annealing,liu2017investigation} and its stoichiometry is shown in the XPS (see SI). The grown 3C-SiC layer, seen in Fig.~\ref{figure1}a, is approximately 10~nm thick and not atomically smooth nor perfectly homogenous with noticeable shadow contrasting around the interface. This is due to changes in material density that occur though the sample lamella ($\sim$100~nm thick) and strain from localized and extended defects. Despite this, the crystallinity in both the grown layer and substrate is evident, as seen in the inset which is the calculated FFT of the TEM image Fig.~\ref{figure1}a. Importantly, no major strain-relieving threading dislocations are observed to propagate into the bulk on either side of the interface. Fig.~\ref{figure1}c is the enlarged area highlighted in Fig.~\ref{figure1}a with a dashed box where high coherence between 3C-SiC and diamond is seen to cover several nanometers. Despite the coherence, the two materials are very slightly misoriented by approximately 1$^\circ$, as shown in orange, comparable to the literature values of 0.52$^\circ$ when diamond was seeded on a 3C-SiC substrate \cite{yaita2017influence}. Coherence over 5-10 nanometres without extended defects in the grown layer indicate that a periodic bonding structure exists at the interface. 

\begin{figure*}[t]
    \centering
    \includegraphics[width=15cm]{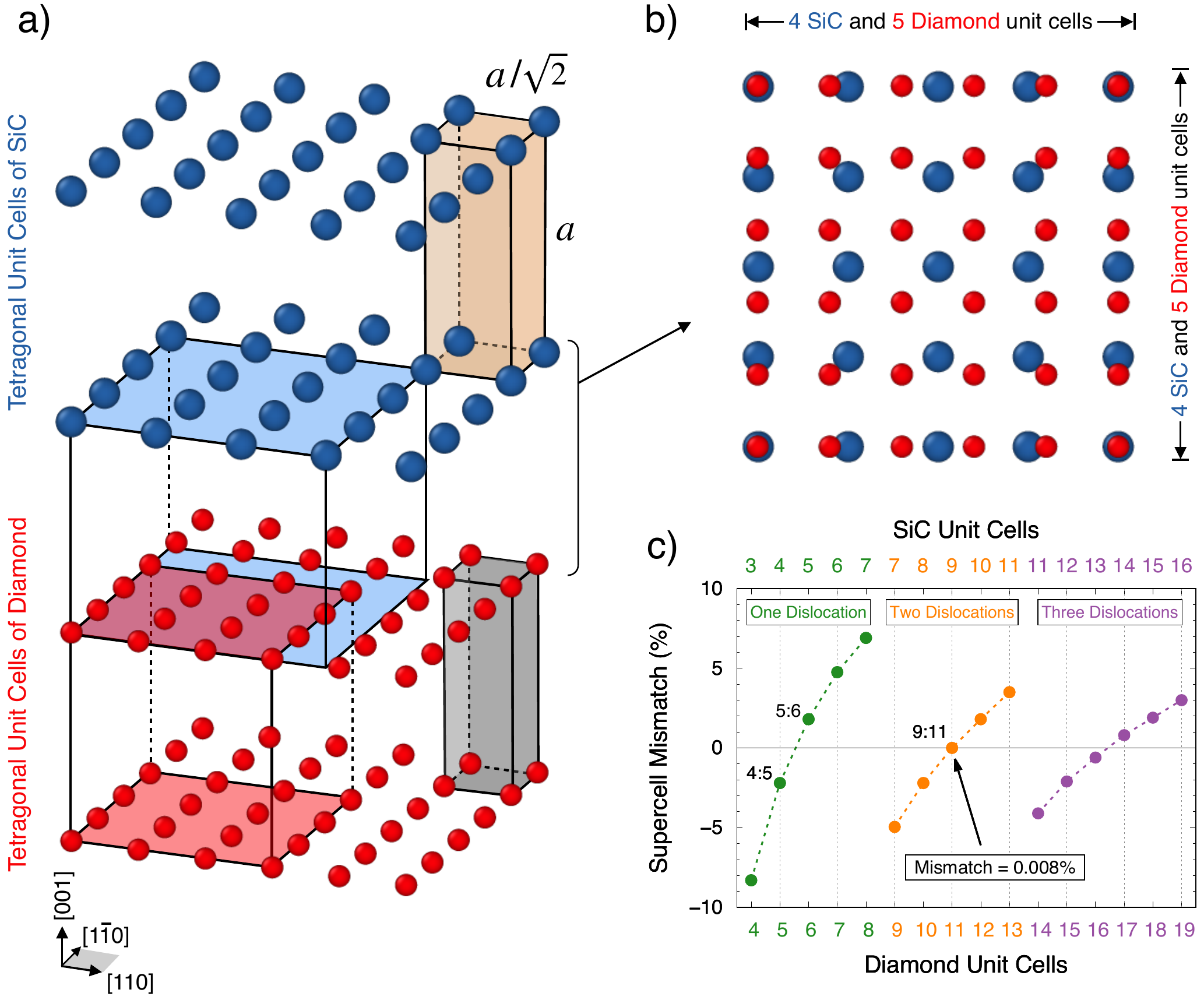}
    \caption{(a) Schematic of four interfacial layers, where each sphere represents a tetragonal unit cell with a 4-atom basis. Shaded areas compare $3\times3$ unit supercells, showing large deviations between 3C-SiC and diamond. (b) Strained $4\times4$ SiC supercell overlaid on a $5\times5$ diamond supercell resulting in a Moir\'e pattern. (c) Plot of supercell mismatch percentage between various combinations of SiC and diamond unit cells. Ratios denote the number of SiC and diamond unit cells, respectively. }
    \label{figure2}
\end{figure*}

The quality of the interface is further demonstrated by the FFT shown in Fig.~\ref{figure1}d, which is calculated from Fig.~\ref{figure1}c. Here, the bright diffraction signals are arranged into inner and outer hexagons which are calculated from the delineated areas in Fig.~\ref{figure1}c; these are caused by 3C-SiC and diamond, respectively. From these FFTs it is clear that the misalignment between the two materials, if any, is very slight with $[1\bar{1}\bar{1}]$, $[002]$ and $[1\bar{1}1]$ signals showing a strong linear dependency, thus providing strong evidence that the grown 3C-SiC used the diamond as a template for crystallization. From these diffraction points, the lattice constants for 3C-SiC and diamond are derived to be 4.42~\AA\ and 3.61~\AA, respectively. The signal peaks were located by fitting Gaussians to intensity line cuts that went through opposing $[002]$ diffraction points to achieve sub-pixel accuracy. The calculated values are within 2\% accuracy of the literature lattice constant values of $a_{\text{SiC}} = 4.360$~\AA\ \cite{Taylor-1960, Jarrendahl-1998} and $a_{\text{diamond}} = 3.567$~\AA\ \cite{kruger-2010} and are well within instrumental error margins. This, in conjunction with the FFT plots, indicate that there is minimal strain between the materials despite a lattice mismatch of $22\%$. Outside the area of high quality crystallinity, however, stacking faults and defects begin to appear in the grown 3C-SiC layer as shown by fault lines in Fig.~\ref{figure1}d which are barely evident in Fig.~\ref{figure1}e (from the ($1\bar{1}\bar{1}$ point to the (002) point in the SiC Fourier transform). The formation of low-strain SiC locally on single-crystal diamond provides strong evidence that this fabrication procedure has promise in bypassing the polycrystalline limitations of diamond growth seen in other diamond heterostructures \cite{koizumi2018power}. 

To understand the origin of the low-strain epitaxial growth in Fig.~\ref{figure1}, molecular dynamics (MD) simulations were performed to establish how strain could be resolved at the interface without propagation of dislocations in SiC. The first step is the determination of the number of unit cells, as summarized schematically in Fig.~\ref{figure2}. Panel (a) shows four layers at the interface where each sphere represents a 4-atom tetragonal unit cell with height $a$ and in-plane width of $a/\sqrt{2}$. The blue and red shading highlights the large (22\%) in-plane mismatch between 3$\times$3 supercells of SiC and diamond. It is visually apparent in Fig.~\ref{figure2}a that 4 unit cells of SiC match reasonably closely to 5 unit cells of diamond, and here the supercell mismatch is just 2.3\%. Aligning the two interfaces for this ratio of 4:5 unit cells results in the Moir\'e pattern shown in Fig.~\ref{figure2}b. From an atomic perspective, the Moir\'e pattern raises a challenge as it is unclear how to reconstruct the interface without creating dangling bonds and other high energy configurations. Studies of lattice mismatch sometimes trivialize these interfaces with one-dimensional schematics that do not account for the more difficult two-dimensional problem. Fig.~\ref{figure2}c shows that the 4:5 combination is not the only possibility for the SiC/diamond interface, since adding an extra unit cell to both sides of the interface yields a 5:6 ratio with a slightly lower supercell mismatch of 1.8\%. More interesting is a 9:11 ratio for which the mismatch falls to an astonishingly low 0.008\%. The 9:11 ratio has previously been noted when patches of coherent diamond grew on 3C-SiC \cite{koizumi2018power}, but no quantitative bonding arrangement was provided. This configuration, which involves two dislocations as indicated by the orange text in Fig.~\ref{figure2}c, is the focus of our attention for the remainder of this manuscript.

\begin{figure}[H]
    \centering
    \includegraphics[width = 0.78\textwidth]{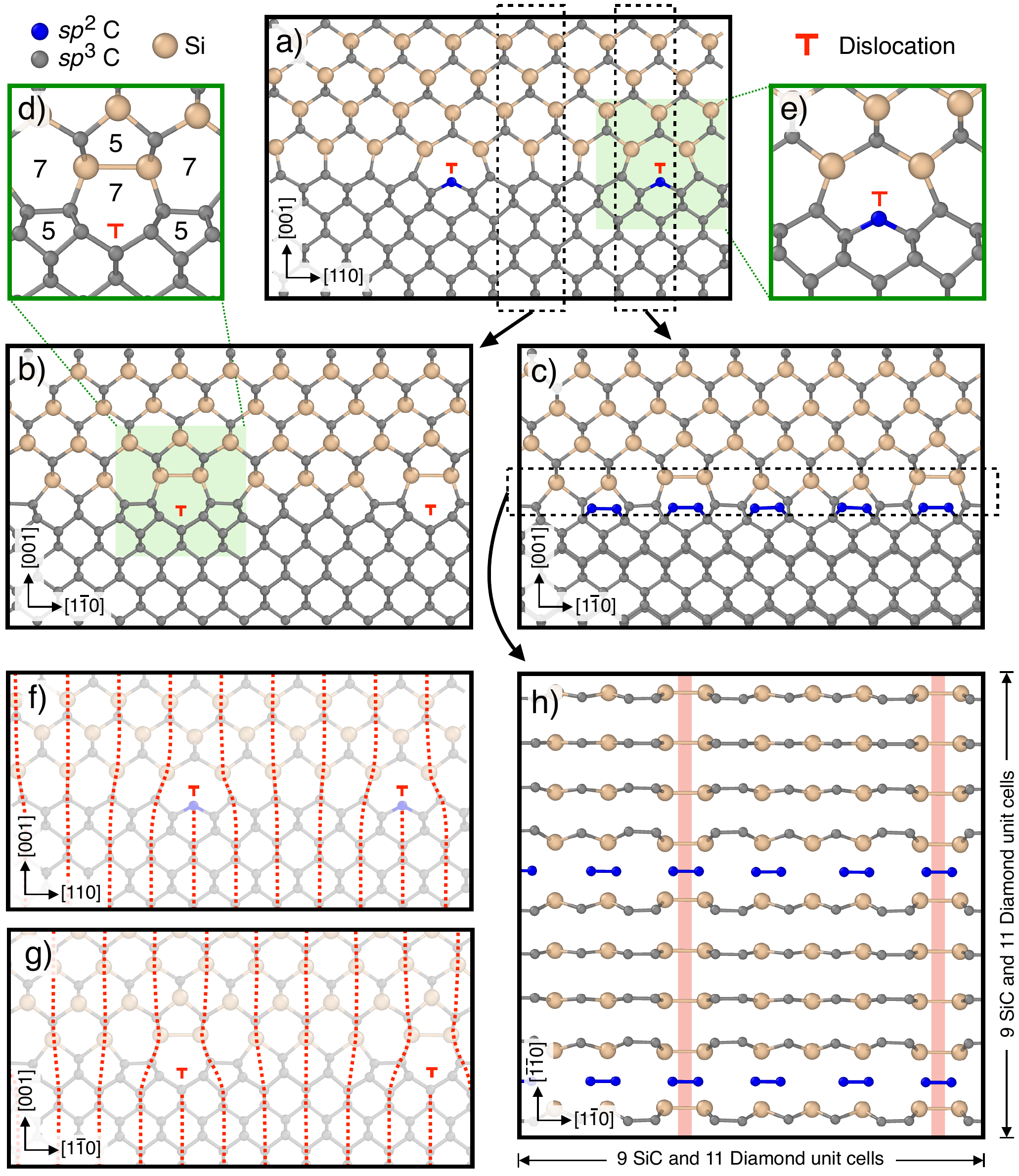}
    \caption{Structure of the reconstructed SiC/diamond interface. (a) View looking along the $[1\bar{1}0]$ direction. Carbon point dislocations (red) correspond with $sp^2$ bonded atoms (blue) every five or six diamond units. Dashed boxes are lattice slices that are viewed down the $[110]$ direction in (b) and (c). Magnified view of shaded green regions are shown in (d,e). Point dislocations in (a) and (b) are highlighted via dashed guidelines in (f) and (g). Dashed box in (c) is a lattice slice shown in plan view in (h). Peach lines indicate columns of Si-Si covalent bonds. }
    \label{figure3}
\end{figure}

The second step in the simulation involves using the Tersoff interatomic potential \cite{Tersoff-PRB-1989} and the LAMMPS package \cite{Plimpton1995} to explore possible interface reconstructions for the 9:11 ratio. This involves: (i) an energy minimization to relax the initial structure, (ii) annealing at a fixed temperature, and (iii) minimization back to zero Kelvin. Four annealing temperatures were used (500, 1000, 1500 and 2000~K) and the annealing time was 20~ps. The best results were achieved at 1500~K, where the interface spontaneously reconstructed to form Si-Si and C-C dimers in two characteristic motifs. One motif spanned 4 SiC unit cells and 5 diamond unit cells, while the other motif spanned 5 SiC unit cells and 6 diamond unit cells. Notably, both of these ratios correspond to small supercell mismatches as seen in Fig.~\ref{figure2}.

In the third step, we manually assemble the 4:5 motif adjacent to a 5:6 motif to create a  reconstruction for the 9:11 ratio, yielding two dislocations. In the transverse direction the annealing simulations reveal additional dislocations every 4 or 5 SiC unit cells. This arrangement is also manually assembled. Next we remove two-fold coordinated ($sp$-bonded) carbons created by the second set of dislocations. The annealing simulations reveal these can be reconstructed in a 2$\times$1 manner analogous to diamond and silicon (001) surfaces, thereby forming $sp^2$ bonds. Since a 2$\times$1 arrangement requires an even number of diamond unit cells, we duplicate the system along one axis to achieve an 18:22 ratio, followed by an energy minimization. As this doubling is a minor detail, all further discussion displays only a single 9:11 section.

Diagrams of the 9:11 interface are illustrated in Fig.~\ref{figure3}. Panel (a) shows that the carbon point dislocations are separated by either five or six diamond unit cells in the $[110]$ direction and are terminated with $sp^2$ bonded carbon (blue). The dashed boxes in panel (a) form slices down the $[1\bar{1}0]$ direction that are seen in Fig.~\ref{figure3}b and \ref{figure3}c. Panel (b) displays the interface without $sp^2$ carbon present; the $sp^2$ carbon seen in Fig.~\ref{figure3}c is mostly isolated from the remainder of the interface as seen in panel (a). Along the $[1\Bar{1}0]$ direction an alternative solution to the surface reconstruction is required. To accommodate misfit strain, single Si-Si covalent bonds form every 4 or 5 SiC unit cells at the SiC interface and single C-C covalent bonds form in cells adjacent to each Si-Si covalent bond in the diamond lattice. This combination of point defects in the $[110]$ and $[1\bar{1}0]$ direction allow the additional lattice units in diamond to terminate at the interface without creating extended defects in the grown film; see the guidelines in Fig.~\ref{figure3}f and g. As a result, defects in the heterostructure are restricted to the two-dimensional plane of the interface, as opposed to propagating through the third-dimension of the grown film. 

The point defect highlighted in Fig.~\ref{figure3}b is magnified in panel (d), demonstrating how the point dislocation involves heptagonal and pentagonal rings to accommodate the extra diamond unit cell. This creates a symmetry around the Si-Si covalent bond which is present elsewhere in the interface, as seen in the other dislocation at the right of Fig.~\ref{figure3}b. Similarly, Fig.~\ref{figure3}e expands the shaded area in panel (a) to emphasise the isolation of the $sp^2$ carbon bond that acts as point dislocation for the epitaxial layer looking along the $[1\bar{1}0]$ direction. In Fig.~\ref{figure3}h, the carbon and silicon atoms at the interface are displayed in plan-view, showing that the $sp^2$ carbons (blue) do not affect the bonding scheme of the remaining interface and that the point dislocation is self-contained within its own row. The peach lines highlight columns of Si-Si covalent bonds, showing how the Si atoms are shifted together slightly while the adjacent carbon atoms are shifted outwards, forming C-C covalent bonds with their neighbouring carbon atom. 

\begin{figure}[H]
    \centering
    \includegraphics[width = 0.48\textwidth]{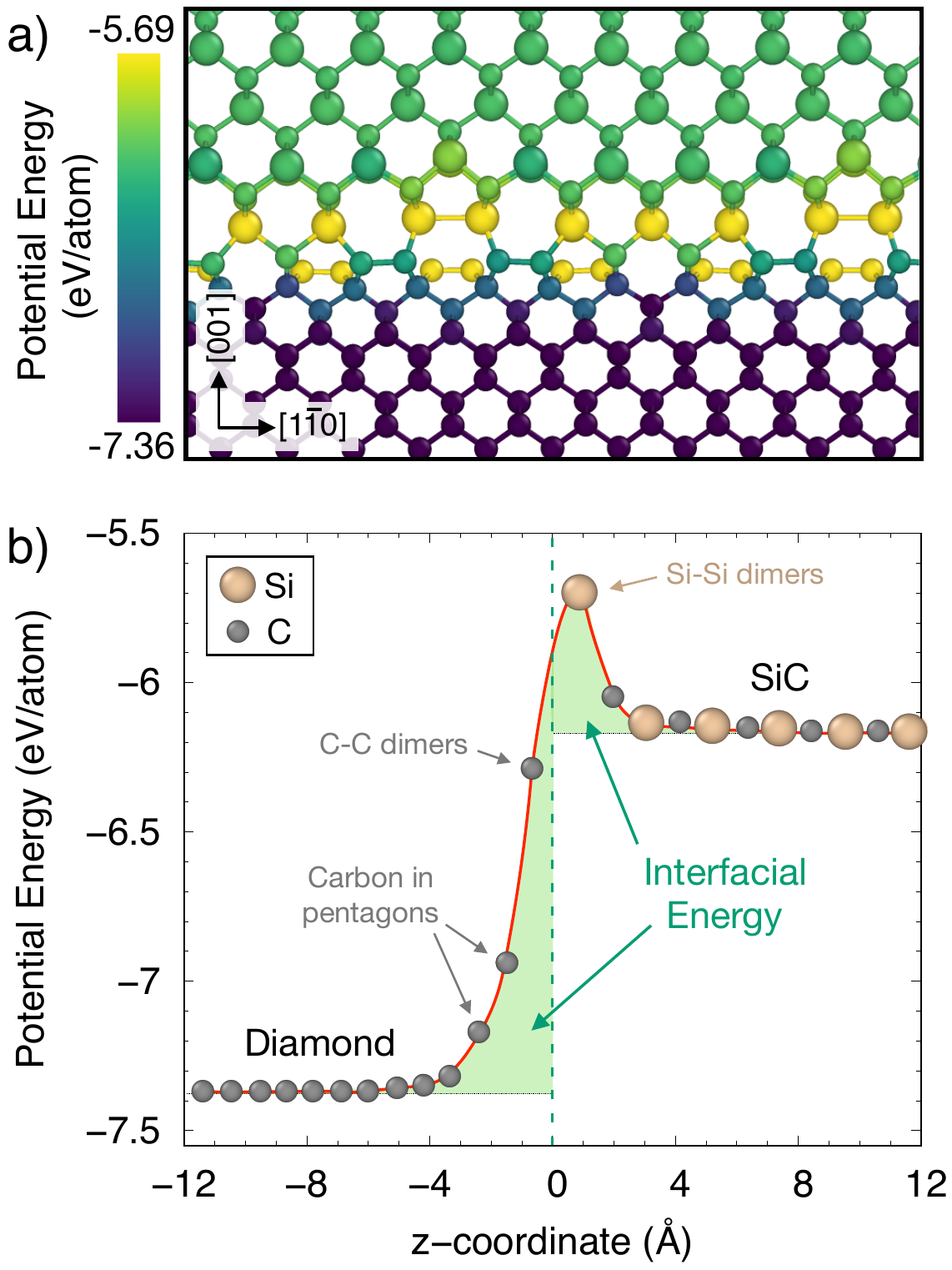}
    \caption{(a) SiC/diamond interface with atoms coloured by their potential energy per atom. (b) Layer-averaged potential energy as a function of distance from the interface. Green shading indicates contributions to the interfacial energy. }
    \label{figure4}
\end{figure}

To quantify the energetics of the reconstruction, the total potential energy was decomposed into atomic contributions and visualised (Fig.~\ref{figure4}a). Away from the interface the atoms adopt the bulk values for diamond and SiC; these compare well to their experimental values of $-$7.37 eV/atom \cite{yin1981ground} and $-$6.34 eV/atom \cite{dyson1996extension,erhart2005analytical}, respectively. The potential energy gradient seen in Fig.~\ref{figure4}a shows that strain is  confined to the interfacial region. The atoms with the most positive potential energy are the interfacial Si atoms ($-5.70$ eV/atom) and the $\pi$-bonded carbon atoms ($-6.28$ eV/atom). However, within a few atomic layers of the interface the potential energy returns to the bulk values, indicative of zero strain. This effect is quantified in Fig.~\ref{figure4}b which plots the layer-averaged potential energy as a function of distance from the interface. On the SiC side only two layers deviate significantly from the bulk, showing that strain is  quickly removed as distance from the interface increases. On the diamond side the strain field penetrates slightly further, reflecting the high stiffness of diamond and non-ideal structures such as pentagonal and heptagonal rings (Fig.~\ref{figure3}d).
Summing the energy cost represented by the green shading in Fig.~\ref{figure4}b yields a relatively modest interfacial energy of 2.87~J/m$^2$. For comparison, the surface energy of 2$\times$1 reconstructed diamond with the Tersoff potential is 6.20~J/m$^2$, while the corresponding value for the Si-rich surface of SiC is 4.01~J/m$^2$. This demonstrates that the 9:11 interface provides a low-strain connection between the two lattice types.

Based on the experimental evidence seen in Fig. \ref{figure1}c, which shows an interface with low strain and high coherence, the 9:11 surface reconstruction demonstrated in Fig. \ref{figure3}f becomes the most likely interface for the region of high crystalline quality as it resolves the surprising coherence and strain issues. Due to the inhomogeneity at the interface with the fabrication procedure, however, it would be challenging for the uniform periodic structure to form across the entire interface. This is partially accommodated for by the ability of carbon to form $\pi$ and $\sigma$ bonds which resolves instances of high strain, but will result in extended defects in larger regions, as seen in Fig. \ref{figure1}a. If greater control over carbon bond manufacturing at the interface was available, not only 3C-SiC, but other materials with large lattice mismatches could form low strain and highly coherent heterostructures with diamond.


In this paper we have shown both theoretically and experimentally that it is possible to form an atomically sharp and minimally strained 3C-SiC film on a single crystal diamond substrate in localized regions. Simulations indicate that this is achieved with an optimal supercell ratio of 9:11 that removes dangling bonds and requires two rows of point dislocations in the $[1\bar{1}0]$ and $[110]$ direction. This is further corroborated through strain calculations that demonstrate that the misfit strain is relieved within five monolayers of the interface. Thus, the high quality interface observed in the experiment is a likely realization of the mechanism suggested by this simulation in localized areas despite the auto-nucleation at the interface. Such a high crystalline quality interface is expected to allow for a minimal thermal barrier junction, enabling improved device performance for high temperature electronic applications.

\section{Methodology}

A single type Ib $\langle 100 \rangle$ crystal with a thin ($<1$~$\mu$m) type IIb diamond overlayer was inserted into vacuum after a short diamond growth which ensured the sample had a clean, high-quality, hydrogen-terminated diamond surface. The single crystal $\langle 100 \rangle$ diamond substrate was transferred from the Melbourne Centre of Nanofabrication to the Soft X-ray Spectroscopy beamline, located at the Australian Synchrotron. Here, it was annealed at 450$^\circ$C for 2 hours to remove possible surface contaminants and adsorbates under UHV. To avoid contamination, the entire growth, LEED, and XPS experiments were performed without removing the sample from UHV conditions. The sample was then flash annealed to 1000$^\circ$C in UHV to remove the hydrogen termination and reconstruct a clean $2\times1$ carbon surface, confirmed to be free of oxygen via XPS. A monolayer of Si was then evaporated onto the substrate and the sample was again flash annealed to 1000$^\circ$C to form a $3\times1$ Si-terminated diamond surface as confirmed by LEED \cite{schenk2015formation}. Approximately 5~nm of Si was then thermally evaporated onto the Si-terminated diamond at room temperature over a period of 9 hours and 25 minutes. 
After the deposition the sample was annealed to 1350$^\circ$C and the formation of SiC was verified via NEXAFS and XPS. High-resolution Transmission Electron Microscopy was conducted \emph{ex-situ} to further study the resultant SiC-on-diamond heterostructure.

The molecular dynamics simulations were performed using the LAMMPS package \cite{Plimpton1995}. Interactions were described using the Tersoff potential for silicon-carbide \cite{Tersoff-PRB-1989}, which is based on earlier Tersoff potentials for silicon \cite{Tersoff-PRL-1986} and carbon \cite{Tersoff-PRL-1988}. With this potential all interactions are captured in a single framework. The interface was rectangular in shape, with the SiC side containing 18 SiC unit-cells in the $[1\bar{1}0]$ direction, 9 SiC unit cells in the $[110]$ direction and 9 SiC unit-cells in the $[001]$ direction. On the other side of the interface, the corresponding number of diamond unit cells was 22, 11 and 11, respectively. Periodic boundary conditions were employed in the $[110]$ and $[1\bar{1}0]$ directions but not in the transverse direction; the system dimensions were approximately $54~\text{\AA}\times 27~\text{\AA}\times40~\text{\AA}$. No attempt was made to manually reconstruct the interface, and the initial coordinates simply consisted of a cleaved SiC crystal adjacent to cleaved diamond crystal. Temperature control was achieved using the Bussi thermostat \cite{Bussi} and simulations were performed in an NVT ensemble (constant number of particles, volume and temperature) with a timestep of 0.2~fs. Coordination analysis involved counting the number of neighbours within cutoffs of 1.85, 2.35 and 2.8~\AA\ for C-C, C-Si and Si-Si distances, respectively. For the purposes of analysis and visualization, atoms are considered to be $sp^2$ and $sp^3$ hybridized if they had three and four neighbours, respectively. Visualization was performed using the OVITO software \cite{OVITO}.

\begin{acknowledgement}

This work was undertaken on the soft x-ray beamline at the Australian Synchrotron, part of ANSTO and performed in part at the Melbourne Centre for Nanofabrication (MCN) in the Victorian Node of the Australian National Fabrication Facility (ANFF). Additionally, TEM was conducted at the Bio21 Advanced Microscopy Facility (the University of Melbourne). Computational resources were also provided by the Pawsey Supercomputing Centre with funding from the Australian Government and the Government of Western Australia. This research was funded by the Australian Research Council Centre of Excellence for Quantum Computation and Communication Technology (No. CE170100012).

\end{acknowledgement}


\providecommand{\latin}[1]{#1}
\makeatletter
\providecommand{\doi}
{\begingroup\let\do\@makeother\dospecials
	\catcode`\{=1 \catcode`\}=2 \doi@aux}
\providecommand{\doi@aux}[1]{\endgroup\texttt{#1}}
\makeatother
\providecommand*\mcitethebibliography{\thebibliography}
\csname @ifundefined\endcsname{endmcitethebibliography}
{\let\endmcitethebibliography\endthebibliography}{}



\begin{mcitethebibliography}{47}
	\providecommand*\natexlab[1]{#1}
	\providecommand*\mciteSetBstSublistMode[1]{}
	\providecommand*\mciteSetBstMaxWidthForm[2]{}
	\providecommand*\mciteBstWouldAddEndPuncttrue
	{\def\EndOfBibitem{\unskip.}}
	\providecommand*\mciteBstWouldAddEndPunctfalse
	{\let\EndOfBibitem\relax}
	\providecommand*\mciteSetBstMidEndSepPunct[3]{}
	\providecommand*\mciteSetBstSublistLabelBeginEnd[3]{}
	\providecommand*\EndOfBibitem{}
	\mciteSetBstSublistMode{f}
	\mciteSetBstMaxWidthForm{subitem}{(\alph{mcitesubitemcount})}
	\mciteSetBstSublistLabelBeginEnd
	{\mcitemaxwidthsubitemform\space}
	{\relax}
	{\relax}
	
	\bibitem[Isberg \latin{et~al.}(2002)Isberg, Hammersberg, Johansson,
	Wikstr{\"o}m, Twitchen, Whitehead, Coe, and Scarsbrook]{isberg2002high}
	Isberg,~J.; Hammersberg,~J.; Johansson,~E.; Wikstr{\"o}m,~T.; Twitchen,~D.~J.;
	Whitehead,~A.~J.; Coe,~S.~E.; Scarsbrook,~G.~A. \emph{Science} \textbf{2002},
	\emph{297}, 1670--1672\relax
	\mciteBstWouldAddEndPuncttrue
	\mciteSetBstMidEndSepPunct{\mcitedefaultmidpunct}
	{\mcitedefaultendpunct}{\mcitedefaultseppunct}\relax
	\EndOfBibitem
	\bibitem[Wort and Balmer(2008)Wort, and Balmer]{wort2008diamond}
	Wort,~C. J.~H.; Balmer,~R.~S. \emph{Materials Today} \textbf{2008}, \emph{11},
	22--28\relax
	\mciteBstWouldAddEndPuncttrue
	\mciteSetBstMidEndSepPunct{\mcitedefaultmidpunct}
	{\mcitedefaultendpunct}{\mcitedefaultseppunct}\relax
	\EndOfBibitem
	\bibitem[Willander \latin{et~al.}(2006)Willander, Friesel, Wahab, and
	Straumal]{willander2006silicon}
	Willander,~M.; Friesel,~M.; Wahab,~Q.-U.; Straumal,~B. \emph{Journal of
		Materials Science: Materials in Electronics} \textbf{2006}, \emph{17},
	1\relax
	\mciteBstWouldAddEndPuncttrue
	\mciteSetBstMidEndSepPunct{\mcitedefaultmidpunct}
	{\mcitedefaultendpunct}{\mcitedefaultseppunct}\relax
	\EndOfBibitem
	\bibitem[Trew \latin{et~al.}(1991)Trew, Yan, and Mock]{trew1991potential}
	Trew,~R.~J.; Yan,~J.-B.; Mock,~P.~M. \emph{Proceedings of the IEEE}
	\textbf{1991}, \emph{79}, 598--620\relax
	\mciteBstWouldAddEndPuncttrue
	\mciteSetBstMidEndSepPunct{\mcitedefaultmidpunct}
	{\mcitedefaultendpunct}{\mcitedefaultseppunct}\relax
	\EndOfBibitem
	\bibitem[Okushi(2001)]{okushi2001high}
	Okushi,~H. \emph{Diamond and Related Materials} \textbf{2001}, \emph{10},
	281--288\relax
	\mciteBstWouldAddEndPuncttrue
	\mciteSetBstMidEndSepPunct{\mcitedefaultmidpunct}
	{\mcitedefaultendpunct}{\mcitedefaultseppunct}\relax
	\EndOfBibitem
	\bibitem[Meyer \latin{et~al.}(2003)Meyer, Dasgupta, Shaddock, Tucker, Fillion,
	Bronecke, Yorinks, and Kraft]{meyer2003silicon}
	Meyer,~L.; Dasgupta,~S.; Shaddock,~D.; Tucker,~J.; Fillion,~R.; Bronecke,~P.;
	Yorinks,~L.; Kraft,~P. \emph{A silicon-carbide micro-capillary pumped loop
		for cooling high power devices}; Proceedings of Ninteenth Annual IEEE
	Semiconductor Thermal Measurement and Management Symposium, 2003., 2003; pp
	364--368\relax
	\mciteBstWouldAddEndPuncttrue
	\mciteSetBstMidEndSepPunct{\mcitedefaultmidpunct}
	{\mcitedefaultendpunct}{\mcitedefaultseppunct}\relax
	\EndOfBibitem
	\bibitem[Weitzel \latin{et~al.}(1996)Weitzel, Palmour, Carter, Moore,
	Nordquist, Allen, Thero, and Bhatnagar]{weitzel1996silicon}
	Weitzel,~C.~E.; Palmour,~J.~W.; Carter,~C.~H.; Moore,~K.; Nordquist,~K.~K.;
	Allen,~S.; Thero,~C.; Bhatnagar,~M. \emph{IEEE transactions on Electron
		Devices} \textbf{1996}, \emph{43}, 1732--1741\relax
	\mciteBstWouldAddEndPuncttrue
	\mciteSetBstMidEndSepPunct{\mcitedefaultmidpunct}
	{\mcitedefaultendpunct}{\mcitedefaultseppunct}\relax
	\EndOfBibitem
	\bibitem[Chen and Washburn(1996)Chen, and Washburn]{chen1996structural}
	Chen,~Y.; Washburn,~J. \emph{Physical Review Letters} \textbf{1996}, \emph{77},
	4046\relax
	\mciteBstWouldAddEndPuncttrue
	\mciteSetBstMidEndSepPunct{\mcitedefaultmidpunct}
	{\mcitedefaultendpunct}{\mcitedefaultseppunct}\relax
	\EndOfBibitem
	\bibitem[People and Bean(1985)People, and Bean]{people1985calculation}
	People,~R.; Bean,~J.~C. \emph{Applied Physics Letters} \textbf{1985},
	\emph{47}, 322--324\relax
	\mciteBstWouldAddEndPuncttrue
	\mciteSetBstMidEndSepPunct{\mcitedefaultmidpunct}
	{\mcitedefaultendpunct}{\mcitedefaultseppunct}\relax
	\EndOfBibitem
	\bibitem[Bean \latin{et~al.}(1984)Bean, Feldman, Fiory, Nakahara, and
	Robinson]{bean1984ge}
	Bean,~J.~C.; Feldman,~L.~C.; Fiory,~A.~T.; Nakahara,~S.~T.; Robinson,~I.~K.
	\emph{Journal of Vacuum Science \& Technology A: Vacuum, Surfaces, and Films}
	\textbf{1984}, \emph{2}, 436--440\relax
	\mciteBstWouldAddEndPuncttrue
	\mciteSetBstMidEndSepPunct{\mcitedefaultmidpunct}
	{\mcitedefaultendpunct}{\mcitedefaultseppunct}\relax
	\EndOfBibitem
	\bibitem[Freund and Suresh(2004)Freund, and Suresh]{freund2004thin}
	Freund,~L.~B.; Suresh,~S. \emph{Thin {F}ilm {M}aterials: {S}tress, {D}efect
		{F}ormation and {S}urface {E}volution}; Cambridge University Press,
	2004\relax
	\mciteBstWouldAddEndPuncttrue
	\mciteSetBstMidEndSepPunct{\mcitedefaultmidpunct}
	{\mcitedefaultendpunct}{\mcitedefaultseppunct}\relax
	\EndOfBibitem
	\bibitem[Dong \latin{et~al.}(2014)Dong, Wen, and Melnik]{dong2014relative}
	Dong,~H.; Wen,~B.; Melnik,~R. \emph{Scientific reports} \textbf{2014},
	\emph{4}, 7037\relax
	\mciteBstWouldAddEndPuncttrue
	\mciteSetBstMidEndSepPunct{\mcitedefaultmidpunct}
	{\mcitedefaultendpunct}{\mcitedefaultseppunct}\relax
	\EndOfBibitem
	\bibitem[Cervenka \latin{et~al.}(2012)Cervenka, Dontschuk, Ladouceur, Duvall,
	and Prawer]{cervenka2012diamond}
	Cervenka,~J.; Dontschuk,~N.; Ladouceur,~F.; Duvall,~S.~G.; Prawer,~S.
	\emph{Applied Physics Letters} \textbf{2012}, \emph{101}, 051902\relax
	\mciteBstWouldAddEndPuncttrue
	\mciteSetBstMidEndSepPunct{\mcitedefaultmidpunct}
	{\mcitedefaultendpunct}{\mcitedefaultseppunct}\relax
	\EndOfBibitem
	\bibitem[Imura \latin{et~al.}(2011)Imura, Hayakawa, Watanabe, Liao, Koide, and
	Amano]{imura2011demonstration}
	Imura,~M.; Hayakawa,~R.; Watanabe,~E.; Liao,~M.; Koide,~Y.; Amano,~H.
	\emph{{P}hysica {S}tatus {S}olidi (RRL)--Rapid Research Letters}
	\textbf{2011}, \emph{5}, 125--127\relax
	\mciteBstWouldAddEndPuncttrue
	\mciteSetBstMidEndSepPunct{\mcitedefaultmidpunct}
	{\mcitedefaultendpunct}{\mcitedefaultseppunct}\relax
	\EndOfBibitem
	\bibitem[Goyal \latin{et~al.}(2010)Goyal, Subrina, Nika, and
	Balandin]{goyal2010reduced}
	Goyal,~V.; Subrina,~S.; Nika,~D.~L.; Balandin,~A.~A. \emph{Applied Physics
		Letters} \textbf{2010}, \emph{97}, 031904\relax
	\mciteBstWouldAddEndPuncttrue
	\mciteSetBstMidEndSepPunct{\mcitedefaultmidpunct}
	{\mcitedefaultendpunct}{\mcitedefaultseppunct}\relax
	\EndOfBibitem
	\bibitem[Jiang and Jia(1995)Jiang, and Jia]{jiang1995diamond}
	Jiang,~X.; Jia,~C.~L. \emph{Applied physics letters} \textbf{1995}, \emph{67},
	1197--1199\relax
	\mciteBstWouldAddEndPuncttrue
	\mciteSetBstMidEndSepPunct{\mcitedefaultmidpunct}
	{\mcitedefaultendpunct}{\mcitedefaultseppunct}\relax
	\EndOfBibitem
	\bibitem[Khmelnitskiy(2015)]{khmelnitskiy2015prospects}
	Khmelnitskiy,~R.~A. \emph{Physics-Uspekhi} \textbf{2015}, \emph{58}, 134\relax
	\mciteBstWouldAddEndPuncttrue
	\mciteSetBstMidEndSepPunct{\mcitedefaultmidpunct}
	{\mcitedefaultendpunct}{\mcitedefaultseppunct}\relax
	\EndOfBibitem
	\bibitem[Schreck \latin{et~al.}(2014)Schreck, Asmussen, Shikata, Arnault, and
	Fujimori]{schreck2014large}
	Schreck,~M.; Asmussen,~J.; Shikata,~S.; Arnault,~J.-C.; Fujimori,~N. \emph{Mrs
		Bulletin} \textbf{2014}, \emph{39}, 504--510\relax
	\mciteBstWouldAddEndPuncttrue
	\mciteSetBstMidEndSepPunct{\mcitedefaultmidpunct}
	{\mcitedefaultendpunct}{\mcitedefaultseppunct}\relax
	\EndOfBibitem
	\bibitem[Yoder(1996)]{yoder1996wide}
	Yoder,~M.~N. \emph{IEEE Transactions on Electron Devices} \textbf{1996},
	\emph{43}, 1633--1636\relax
	\mciteBstWouldAddEndPuncttrue
	\mciteSetBstMidEndSepPunct{\mcitedefaultmidpunct}
	{\mcitedefaultendpunct}{\mcitedefaultseppunct}\relax
	\EndOfBibitem
	\bibitem[Schreck \latin{et~al.}(2001)Schreck, H{\"o}rmann, Roll, Lindner, and
	Stritzker]{schreck2001diamond}
	Schreck,~M.; H{\"o}rmann,~F.; Roll,~H.; Lindner,~J. K.~N.; Stritzker,~B.
	\emph{Applied Physics Letters} \textbf{2001}, \emph{78}, 192--194\relax
	\mciteBstWouldAddEndPuncttrue
	\mciteSetBstMidEndSepPunct{\mcitedefaultmidpunct}
	{\mcitedefaultendpunct}{\mcitedefaultseppunct}\relax
	\EndOfBibitem
	\bibitem[Koizumi \latin{et~al.}(2018)Koizumi, Umezawa, Pernot, and
	Suzuki]{koizumi2018power}
	Koizumi,~S.; Umezawa,~H.; Pernot,~J.; Suzuki,~M. \emph{Power Electronics Device
		Applications of Diamond Semiconductors}; Woodhead Publishing, 2018\relax
	\mciteBstWouldAddEndPuncttrue
	\mciteSetBstMidEndSepPunct{\mcitedefaultmidpunct}
	{\mcitedefaultendpunct}{\mcitedefaultseppunct}\relax
	\EndOfBibitem
	\bibitem[Aleksov \latin{et~al.}(2003)Aleksov, Kubovic, Kaeb, Spitzberg,
	Bergmaier, Dollinger, Bauer, Schreck, Stritzker, and
	Kohn]{aleksov2003diamond}
	Aleksov,~A.; Kubovic,~M.; Kaeb,~N.; Spitzberg,~U.; Bergmaier,~A.;
	Dollinger,~G.; Bauer,~T.~H.; Schreck,~M.; Stritzker,~B.; Kohn,~E.
	\emph{Diamond and Related Materials} \textbf{2003}, \emph{12}, 391--398\relax
	\mciteBstWouldAddEndPuncttrue
	\mciteSetBstMidEndSepPunct{\mcitedefaultmidpunct}
	{\mcitedefaultendpunct}{\mcitedefaultseppunct}\relax
	\EndOfBibitem
	\bibitem[Stoner and Glass(1992)Stoner, and Glass]{stoner1992textured}
	Stoner,~B.~R.; Glass,~J.~T. \emph{Applied {P}hysics {L}etters} \textbf{1992},
	\emph{60}, 698--700\relax
	\mciteBstWouldAddEndPuncttrue
	\mciteSetBstMidEndSepPunct{\mcitedefaultmidpunct}
	{\mcitedefaultendpunct}{\mcitedefaultseppunct}\relax
	\EndOfBibitem
	\bibitem[Yamada \latin{et~al.}(2014)Yamada, Chayahara, Mokuno, Kato, and
	Shikata]{yamada20142}
	Yamada,~H.; Chayahara,~A.; Mokuno,~Y.; Kato,~Y.; Shikata,~S. \emph{Applied
		Physics Letters} \textbf{2014}, \emph{104}, 102110\relax
	\mciteBstWouldAddEndPuncttrue
	\mciteSetBstMidEndSepPunct{\mcitedefaultmidpunct}
	{\mcitedefaultendpunct}{\mcitedefaultseppunct}\relax
	\EndOfBibitem
	\bibitem[Yamada \latin{et~al.}(2013)Yamada, Chayahara, Mokuno, Tsubouchi, and
	Shikata]{yamada2013uniform}
	Yamada,~H.; Chayahara,~A.; Mokuno,~Y.; Tsubouchi,~N.; Shikata,~S.-I.
	\emph{Diamond and Related Materials} \textbf{2013}, \emph{33}, 27--31\relax
	\mciteBstWouldAddEndPuncttrue
	\mciteSetBstMidEndSepPunct{\mcitedefaultmidpunct}
	{\mcitedefaultendpunct}{\mcitedefaultseppunct}\relax
	\EndOfBibitem
	\bibitem[Hirama \latin{et~al.}(2010)Hirama, Taniyasu, and
	Kasu]{hirama2010heterostructure}
	Hirama,~K.; Taniyasu,~Y.; Kasu,~M. \emph{Journal of Applied Physics}
	\textbf{2010}, \emph{108}, 013528\relax
	\mciteBstWouldAddEndPuncttrue
	\mciteSetBstMidEndSepPunct{\mcitedefaultmidpunct}
	{\mcitedefaultendpunct}{\mcitedefaultseppunct}\relax
	\EndOfBibitem
	\bibitem[Hirama \latin{et~al.}(2011)Hirama, Taniyasu, and
	Kasu]{hirama2011algan}
	Hirama,~K.; Taniyasu,~Y.; Kasu,~M. \emph{Applied Physics Letters}
	\textbf{2011}, \emph{98}, 162112\relax
	\mciteBstWouldAddEndPuncttrue
	\mciteSetBstMidEndSepPunct{\mcitedefaultmidpunct}
	{\mcitedefaultendpunct}{\mcitedefaultseppunct}\relax
	\EndOfBibitem
	\bibitem[Alomari \latin{et~al.}(2010)Alomari, Dussaigne, Martin, Grandjean,
	Gaqui{\`e}re, and Kohn]{alomari2010algan}
	Alomari,~M.; Dussaigne,~A.; Martin,~D.; Grandjean,~N.; Gaqui{\`e}re,~C.;
	Kohn,~E. \emph{Electronics Letters} \textbf{2010}, \emph{46}, 299--301\relax
	\mciteBstWouldAddEndPuncttrue
	\mciteSetBstMidEndSepPunct{\mcitedefaultmidpunct}
	{\mcitedefaultendpunct}{\mcitedefaultseppunct}\relax
	\EndOfBibitem
	\bibitem[Hirama \latin{et~al.}(2012)Hirama, Kasu, and
	Taniyasu]{hirama2012growth}
	Hirama,~K.; Kasu,~M.; Taniyasu,~Y. \emph{Japanese Journal of Applied Physics}
	\textbf{2012}, \emph{51}, 01AG09\relax
	\mciteBstWouldAddEndPuncttrue
	\mciteSetBstMidEndSepPunct{\mcitedefaultmidpunct}
	{\mcitedefaultendpunct}{\mcitedefaultseppunct}\relax
	\EndOfBibitem
	\bibitem[Schenk \latin{et~al.}(2015)Schenk, Tadich, Sear, O'Donnell, Ley,
	Stacey, and Pakes]{schenk2015formation}
	Schenk,~A.~K.; Tadich,~A.; Sear,~M.; O'Donnell,~K.~M.; Ley,~L.; Stacey,~A.;
	Pakes,~C.~I. \emph{Applied Physics Letters} \textbf{2015}, \emph{106},
	191603\relax
	\mciteBstWouldAddEndPuncttrue
	\mciteSetBstMidEndSepPunct{\mcitedefaultmidpunct}
	{\mcitedefaultendpunct}{\mcitedefaultseppunct}\relax
	\EndOfBibitem
	\bibitem[Chang \latin{et~al.}(1999)Chang, Hsieh, Pong, Tsai, Dann, Chien,
	Tseng, Chen, Wei, Chen, \latin{et~al.} others]{chang1999x}
	others,, \latin{et~al.}  \emph{Journal of applied physics} \textbf{1999},
	\emph{86}, 5609--5613\relax
	\mciteBstWouldAddEndPuncttrue
	\mciteSetBstMidEndSepPunct{\mcitedefaultmidpunct}
	{\mcitedefaultendpunct}{\mcitedefaultseppunct}\relax
	\EndOfBibitem
	\bibitem[Prado \latin{et~al.}(2003)Prado, D’addio, Fantini, Pereyra, and
	Flank]{prado2003annealing}
	Prado,~R.~J.; D’addio,~T.~F.; Fantini,~M. C. d.~A.; Pereyra,~I.; Flank,~A.~M.
	\emph{Journal of non-crystalline solids} \textbf{2003}, \emph{330},
	196--215\relax
	\mciteBstWouldAddEndPuncttrue
	\mciteSetBstMidEndSepPunct{\mcitedefaultmidpunct}
	{\mcitedefaultendpunct}{\mcitedefaultseppunct}\relax
	\EndOfBibitem
	\bibitem[Liu \latin{et~al.}(2017)Liu, Yang, Gao, Liu, Huang, Zhou, and
	Sham]{liu2017investigation}
	Liu,~M.; Yang,~X.; Gao,~Y.; Liu,~R.; Huang,~H.; Zhou,~X.; Sham,~T.~K.
	\emph{Journal of the European Ceramic Society} \textbf{2017}, \emph{37},
	1253--1259\relax
	\mciteBstWouldAddEndPuncttrue
	\mciteSetBstMidEndSepPunct{\mcitedefaultmidpunct}
	{\mcitedefaultendpunct}{\mcitedefaultseppunct}\relax
	\EndOfBibitem
	\bibitem[Yaita \latin{et~al.}(2017)Yaita, Natal, Saddow, Hatano, and
	Iwasaki]{yaita2017influence}
	Yaita,~J.; Natal,~M.; Saddow,~S.~E.; Hatano,~M.; Iwasaki,~T. \emph{Applied
		Physics Express} \textbf{2017}, \emph{10}, 045502\relax
	\mciteBstWouldAddEndPuncttrue
	\mciteSetBstMidEndSepPunct{\mcitedefaultmidpunct}
	{\mcitedefaultendpunct}{\mcitedefaultseppunct}\relax
	\EndOfBibitem
	\bibitem[Taylor and Jones(1960)Taylor, and Jones]{Taylor-1960}
	Taylor,~A.; Jones,~R.~M. In \emph{Silicon Carbide -- A High Temperature
		Semiconductor}; O'Connor,~J.~R., Smiltens,~J., Eds.; Pergamon Press, 1960; pp
	147--156\relax
	\mciteBstWouldAddEndPuncttrue
	\mciteSetBstMidEndSepPunct{\mcitedefaultmidpunct}
	{\mcitedefaultendpunct}{\mcitedefaultseppunct}\relax
	\EndOfBibitem
	\bibitem[Järrendahl and Davis(1998)Järrendahl, and Davis]{Jarrendahl-1998}
	Järrendahl,~K.; Davis,~R.~F. In \emph{SiC Materials and Devices}; Park,~Y.~S.,
	Ed.; Semiconductors and Semimetals; Elsevier, 1998; Vol.~52; pp 1 -- 20\relax
	\mciteBstWouldAddEndPuncttrue
	\mciteSetBstMidEndSepPunct{\mcitedefaultmidpunct}
	{\mcitedefaultendpunct}{\mcitedefaultseppunct}\relax
	\EndOfBibitem
	\bibitem[Kr{\"u}ger(2010)]{kruger-2010}
	Kr{\"u}ger,~A. \emph{Carbon Materials and Nanotechnology}; Wiley, 2010\relax
	\mciteBstWouldAddEndPuncttrue
	\mciteSetBstMidEndSepPunct{\mcitedefaultmidpunct}
	{\mcitedefaultendpunct}{\mcitedefaultseppunct}\relax
	\EndOfBibitem
	\bibitem[Tersoff(1989)]{Tersoff-PRB-1989}
	Tersoff,~J. \emph{Physical Review B} \textbf{1989}, \emph{39}, 5566--5568\relax
	\mciteBstWouldAddEndPuncttrue
	\mciteSetBstMidEndSepPunct{\mcitedefaultmidpunct}
	{\mcitedefaultendpunct}{\mcitedefaultseppunct}\relax
	\EndOfBibitem
	\bibitem[Plimpton(1995)]{Plimpton1995}
	Plimpton,~S. \emph{J. Chem. Phys.} \textbf{1995}, \emph{117}, 1--19\relax
	\mciteBstWouldAddEndPuncttrue
	\mciteSetBstMidEndSepPunct{\mcitedefaultmidpunct}
	{\mcitedefaultendpunct}{\mcitedefaultseppunct}\relax
	\EndOfBibitem
	\bibitem[Yin and Cohen(1981)Yin, and Cohen]{yin1981ground}
	Yin,~M.~T.; Cohen,~M.~L. \emph{Physical Review B} \textbf{1981}, \emph{24},
	6121\relax
	\mciteBstWouldAddEndPuncttrue
	\mciteSetBstMidEndSepPunct{\mcitedefaultmidpunct}
	{\mcitedefaultendpunct}{\mcitedefaultseppunct}\relax
	\EndOfBibitem
	\bibitem[Dyson and Smith(1996)Dyson, and Smith]{dyson1996extension}
	Dyson,~A.~J.; Smith,~P.~V. \emph{Surface science} \textbf{1996}, \emph{355},
	140--150\relax
	\mciteBstWouldAddEndPuncttrue
	\mciteSetBstMidEndSepPunct{\mcitedefaultmidpunct}
	{\mcitedefaultendpunct}{\mcitedefaultseppunct}\relax
	\EndOfBibitem
	\bibitem[Erhart and Albe(2005)Erhart, and Albe]{erhart2005analytical}
	Erhart,~P.; Albe,~K. \emph{Physical Review B} \textbf{2005}, \emph{71},
	035211\relax
	\mciteBstWouldAddEndPuncttrue
	\mciteSetBstMidEndSepPunct{\mcitedefaultmidpunct}
	{\mcitedefaultendpunct}{\mcitedefaultseppunct}\relax
	\EndOfBibitem
	\bibitem[Tersoff(1986)]{Tersoff-PRL-1986}
	Tersoff,~J. \emph{Physical Review Letters} \textbf{1986}, \emph{56},
	632--635\relax
	\mciteBstWouldAddEndPuncttrue
	\mciteSetBstMidEndSepPunct{\mcitedefaultmidpunct}
	{\mcitedefaultendpunct}{\mcitedefaultseppunct}\relax
	\EndOfBibitem
	\bibitem[Tersoff(1988)]{Tersoff-PRL-1988}
	Tersoff,~J. \emph{Physical Review Letters} \textbf{1988}, \emph{61},
	2879--2882\relax
	\mciteBstWouldAddEndPuncttrue
	\mciteSetBstMidEndSepPunct{\mcitedefaultmidpunct}
	{\mcitedefaultendpunct}{\mcitedefaultseppunct}\relax
	\EndOfBibitem
	\bibitem[Bussi \latin{et~al.}(2007)Bussi, Donadio, and Parrinello]{Bussi}
	Bussi,~G.; Donadio,~D.; Parrinello,~M. \emph{The Journal of Chemical Physics}
	\textbf{2007}, \emph{126}, 014101\relax
	\mciteBstWouldAddEndPuncttrue
	\mciteSetBstMidEndSepPunct{\mcitedefaultmidpunct}
	{\mcitedefaultendpunct}{\mcitedefaultseppunct}\relax
	\EndOfBibitem
	\bibitem[Stukowski(2010)]{OVITO}
	Stukowski,~A. \emph{Modell. Simul. Mater. Sci. Eng.} \textbf{2010}, \emph{18},
	015012\relax
	\mciteBstWouldAddEndPuncttrue
	\mciteSetBstMidEndSepPunct{\mcitedefaultmidpunct}
	{\mcitedefaultendpunct}{\mcitedefaultseppunct}\relax
	\EndOfBibitem
\end{mcitethebibliography}
\end{document}